\documentclass[12pt,a4paper]{article}

\RequirePackage[l2tabu, orthodox]{nag}

\usepackage{mjheppub}
\usepackage{color}
\usepackage{lipsum}
\usepackage{amsthm,amssymb,amsmath,epic,eepic,float}
\usepackage{rotating,epsfig,indentfirst,array,varioref}
\usepackage{appendix,marginnote,tikz,pgf,mathtools}
\usepackage{setspace}

\usepackage{tikz}
\usetikzlibrary{decorations.pathreplacing}
\usetikzlibrary{calc}

\usepackage{color}

\newcommand{\dif}{\mathrm{d}}

\newcommand{\ZZ}{\mathbb{Z}}

\newtheorem{prop}{Proposition}[section]

\newdimen\tableauside\tableauside=1.0ex
\newdimen\tableaurule\tableaurule=0.4pt
\newdimen\tableaustep
\def\phantomhrule#1{\hbox{\vbox to0pt{\hrule height\tableaurule
width#1\vss}}}
\def\phantomvrule#1{\vbox{\hbox to0pt{\vrule width\tableaurule
height#1\hss}}}
\def\sqr{\vbox{%
  \phantomhrule\tableaustep

\hbox{\phantomvrule\tableaustep\kern\tableaustep\phantomvrule\tableaustep}%
  \hbox{\vbox{\phantomhrule\tableauside}\kern-\tableaurule}}}
\def\squares#1{\hbox{\count0=#1\noindent\loop\sqr
  \advance\count0 by-1 \ifnum\count0>0\repeat}}
\def\tableau#1{\vcenter{\offinterlineskip
  \tableaustep=\tableauside\advance\tableaustep by-\tableaurule
  \kern\normallineskip\hbox
    {\kern\normallineskip\vbox
      {\gettableau#1 0 }%
     \kern\normallineskip\kern\tableaurule}%
  \kern\normallineskip\kern\tableaurule}}
\def\gettableau#1 {\ifnum#1=0\let\next=\null\else
  \squares{#1}\let\next=\gettableau\fi\next}

\def\be{\begin{equation}}
\def\ee{\end{equation}}
\def\ba{\begin{array}}
\def\ea{\end{array}}

\restylefloat{figure}

\newcommand{\CC}{\mathbb{C}}
\newcommand{\MM}{\mathcal{M}}

\newcommand{\FF}{\mathcal{F}}
\newcommand{\UU}{\mathbb{U}}
\newcommand{\WW}{\mathbb{W}}

\def\Res{\operatorname{Res}}

\renewcommand{\mod}{\textup{mod}\,}

\begin{document}

\title{\boldmath 
On the construction of the correlation numbers in Minimal Liouville Gravity 
} 
\author{Konstantin Aleshkin   $^1$,}
\author{Vladimir Belavin  $^2$}

\affiliation{$^1$ L.D. Landau Institute for Theoretical Physics, 
 Akademika Semenova av., 1-A, \\Chernogolovka, 142432  Moscow region, Russia,
 \newline
 \hspace*{1.5mm} International School of Advanced Studies (SISSA), via Bonomea 265, 34136 Trieste, Italy}
\affiliation{\vspace{2mm} $^2$ 
 I.E. Tamm Department of Theoretical Physics, P.N. Lebedev Physical Institute,\\
Leninsky Avenue 53, 119991 Moscow, Russia,
\newline
 \hspace*{1.5mm} Department of Quantum Physics, Institute for Information Transmission Problems,\\
Bolshoy Karetny per. 19, 127994 Moscow, Russia,
\newline
  \hspace*{1.5mm} Moscow Institute of Physics and Technology, 
Dolgoprudnyi, 141700 Moscow region, Russia
}

\emailAdd{kkcnst@gmail.com,
belavin@lpi.ru
}

\abstract{The computation of the correlation numbers in Minimal Liouville Gravity involves an integration over moduli
 spaces of complex curves. There are two independent approaches to the calculation: the direct one, based on the CFT methods
 and  Liouville higher equations of motion, and the alternative one, motivated by discrete description of 2D gravity and based
 on the Douglas string equation. However these two approaches give rise to the results that are not always consistent among
 themselves. In this paper we explore this problem. We show that in order to reconcile two methods the so-called discrete
 terms in the operator product expansion in the underlying Liouville theory must be properly
 taken into account. In this way we propose modified version of the expression for four-point correlation number and find
 full agreement between direct and alternative approaches. Our result allows to consider correlators without any 
restrictions on the number of conformal blocks contributing to the matter sector correlation function. }


\maketitle
\flushbottom

\section{Introduction}
\label{section.01.introduction}
\bigskip

Minimal Liouville Gravity (MLG) represents an interesting solvable model of 2D quantum gravity~\cite{Polyakov1}.
The role of the matter in MLG is performed by Minimal Models (MM) of conformal field theory, while the gravitational sector is described by  Liouville Field Theory (LFT). The observables  are built from the primary fields in both sectors and the definition of the correlation numbers involves an  integration over moduli space of 
n-punctured surface\footnote{In this paper we focus on the correlation numbers on a sphere.}.

The direct approach for evaluating the correlation numbers consists of
 computing MM and
LFT correlation functions, and integrating then their product over the moduli space (see, for example~\cite{Zamol3pt}). 
However, for multi-point correlators this method becomes rather complicated. In~\cite{BZ1} so called
 higher equations of motion (HEM) in LFT~\cite{HEM} were used to reduce the moduli integration in
 four-point correlation numbers to the boundary  terms, however this method has certain restrictions.
In what follows
we will call ``HEM formula'' the formula for four-point correlation numbers obtained in~\cite{BZ1}.
It was derived using an assumption that the correlation function  in the matter sector  involves
 a maximal number of conformal blocks in the conformal block decomposition~\cite{BPZ}. 

Another approach to Minimal Liouville Gravity, based on the Douglas string
 equation\footnote{We do not discuss another approach to two dimensional gravity - cohomological
 field theories.} (DSE)
and motivated by the Matrix Models\footnote{For more details see \cite{GinMoor} and references therein.},
allows to compute correlation numbers using the specific solution of the
Douglas equation~\cite{AVDouglas, unitary, BDM}. The observables in DSE approach have natural scaling properties, which allow to identify them with
the observables in MLG~\cite{KPZ}. Further research was
made in~\cite{MSS} in order to compare the direct and DSE approaches. 
In~\cite{BZ_MM} three- and four-point 
correlation numbers in the Lee-Yang  series $(2,p)$ of Minimal Models  were computed  with the help of resonance transformations and the agreement of 
CFT computation
(when it is applicable) and DSE approach computation  was shown.  However, in~\cite{BDM}, where the 
connection between DSE and Frobenius manifolds was established, it was shown that for $(3, p)$
 series 
of MM  it is impossible to obtain full agreement between direct and
DSE approaches even on the  three-point level. Namely, some of the correlators that should vanish according
 to 
MM fusion rules appear to be non zero in the Douglas equation computation. Later, in~\cite{BelavinRud}
an alternative description\footnote{The difference with the previous one is that this approach
 does not require resonance transformations.}, also based on  the Douglas string equation, was proposed for the Lee-Yang series, that
was consistent with the two previous results in the region where HEM formula is applicable but gave different results with the  DSE computation based on the resonance transformation 
in the region where HEM formula cannot be used because of the constraint on the number of conformal blocks.

Thus, we have the following natural questions. The first one is 
how to correct HEM formula to make it applicable, i.e. consistent with the direct moduli integration,
 without any restrictions.
The second one is which of the two DSE formulae is consistent
 with the original direct approach if any.
The main result of this paper is an answer to both of these questions. Namely, we
 modify HEM formula for Lee-Yang series so that it is consistent with the direct computation and show that
 Douglas equation approach as in~\cite{BZ_MM, BDM} gives the same result.

To generalize HEM formula we notice that in fact it was derived in the case, where the matter sector is
represented by the Generalized Minimal Model (GMM)~\cite{Zamol3pt}. GMM is a certain
modification of the Minimal Models which can have arbitrary central charge $<1$
and fields with arbitrary complex conformal dimensions.
 There is an explicit formula for the three-point
function of this model~\cite{Zamol3pt} which coincides with MM three-point function whenever the
last one is non-zero. However we find out that in this theory some of the correlation numbers
are ill-defined, that becomes clear after analysing analytical structure of GMM 3-point functions
and Liouville OPE discrete terms. Taking limits of well-defined correlators
with all except one non-degenerate values of the conformal dimensions, we are able to find the
 exact answer for the genuine Minimal Models correlator. 

This consideration allows to modify properly  the HEM formula.
 The corrected formula is in full agreement with
all numerical checks and also with the Douglas equation formula, so that it fixes a previous uncertainty 
for the Lee-Yang series.\\

The paper is organized as follows. 
In Section~\ref{sec:prel} we remind some facts from the Minimal Models, GMM
 and Liouville theory and fix some notations. In Section~\ref{sec:lg} we talk about
the Liouville Gravity, construct correlation numbers -- the main object of study in this paper
and recall HEM formula.  In Section~\ref{sec:discrete}
we discuss discrete terms in Liouville OPE. The main results of the paper are presented in
Sections~\ref{sec:GHEM} and~\ref{sec:comp}. Finally, in the section~\ref{sec:disc}
we discuss our results and some further questions. In the appendix we
 describe the direct approach for computing correlation numbers, present
several formulae for convenience and prove some proposition.

\section{Preliminaries} \label{sec:prel}
Here we remind some facts about MLG ingredients, Minimal Models and
Liouville Field Theory, and set our notations.

\paragraph{Ordinary and Generalized Minimal Models.}
Minimal models $\mathcal{M}(p'/p)$ are rational conformal field theories
~\cite{BPZ}. The fields
belong to a sum of a finite number of highest weight representations of Virasoro algebra

\begin{equation}
\lbrack L_n,L_m]=(n-m)L_{n+m}+\frac{c}{12}(n^3-n)\delta_{n,-m}\;,
\label{The algebra}
\end{equation}
with central charge 
\begin{equation*}
c_M = 1 - 6 q^2\;,
\end{equation*}
where the parameter $q$ is given by
\begin{equation*}
q = b^{-1} - b, \; b = \sqrt{p'/p}\;.
\end{equation*}
In this setting $(p'-1)(p-1)$ primary fields $\Phi_{m,n}(x)$ \footnote{In what
 follows we omit dependence on $x$ where it is unnecessary.} correspond to the highest weight 
vectors with weights $L_0 \Phi_{m,n} = \Delta^M_{m,n} \Phi_{m,n}$. For the 
conformal dimensions $\Delta^M_{m,n}$ it is convenient to introduce parameters
\begin{equation*}
\lambda_{m,n} = \frac{m b + n b^{-1}}{2}\;, \;
\alpha_{m,n} = \lambda_{m, -n} - q/2\;,
\end{equation*}
so that 
\begin{multline*}
\Delta^M_{m,n} = \frac{(m b^{-1} - n b)^2 - (b^{-1} - b)^2}{4} = \\
\frac{(m p - n p')^2 - (p - p')^2}{4 p p'} = \alpha_{m,n}(\alpha_{m,n}-q) =
 q^2/4 - \lambda_{m,-n}^2\;.
\end{multline*}

The MM structure constants $\mathbb{C}^k_{ij}$ \footnote{Here $i, j, k$ run through the pairs $(m, n)$.}
can be found in ~\cite{DF1}. The OPE
 satisfies the so-called fusion rules which can
be symbolically represented as
\begin{equation} \label{MMOPE}
\Phi_{m_1, n_1} \Phi_{m_2, n_2} = \sum\limits_{r, s} [\Phi_{r, s}]\;,
\end{equation}
where $[\Phi]$ denotes the contribution of the primary field $\Phi$ and all its descendants 
and $|m_1-m_2|+1 : r
 : \mathrm{min}(m_1+m_2-1,\; 2 p' - m_1 - m_2 - 1), \; |n_1-n_2|+1 : s :
 \mathrm{min}(n_1+n_2-1,\; 2 p - n_1 - n_2 - 1)$. These fusion rules
 are equivalent to the fusion algebra
of integrable $SL(2)_{p, p'}$ representations with $\ZZ/2\ZZ$ identification $(n, m) \to (p'-n, p-m)$.
 The structure constants are zero when these rules
are not satisfied.

For analytic computations in Minimal Liouville Gravity
it is instructive to consider Generalized Minimal Models, that is a modification
of the ordinary MM. The central charge in GMM can
be an arbitrary real number less then one and primary fields of the model
can have arbitrary complex dimensions (see for example~\cite{Zamol3pt} for discussion).
Namely, one introduces primary fields $\Phi_{\alpha}$ with dimension $\Delta^M_{\alpha} = 
\alpha(\alpha-q)$. In particular, one has  $\Delta^M_{m, n} = \Delta^M_{\alpha_{m,n}}$.

With this construction more general structure constants $\mathbb{C}^M(\alpha_1,
\alpha_2, \alpha_3)$ were calculated in~\cite{Zamol3pt}

\begin{equation} \label{mat3pt}
\mathbb{C}^M(\alpha_1,
\alpha_2, \alpha_3) =
A \Upsilon(\alpha + b - q) \prod_i \frac{ \Upsilon(\alpha - 2\alpha_i + b)}
{[\Upsilon(2\alpha_i + b)\Upsilon(2\alpha_i + b - q)]^{1/2}}\;,
\end{equation}
where $\alpha = \sum \alpha_i$ and the normalization factor
\begin{equation}
A = \frac{b^{b^{-2}-b^2 - 1} [\gamma(b^2) \gamma(b^{-2}-1)]^{1/2}}{\Upsilon(b)}\;.
\end{equation}
Here $\gamma(x) = \Gamma(x)/\Gamma(1-x)$
and special function $\Upsilon(x) = \Upsilon_b(x)$ is an entire function of complex domain
with zeros in $x = -n b^{-1} - m b $ and $ (n+1) b^{-1} + (m+1) b$,  where $n, m$
 are non-negative integers  (see for example~\cite{Teschner}).

Perhaps, the most important data one can extract from this expression are zeros
 and poles, which can occur, for example, when the fields are degenerate, that is $\alpha_i = \alpha_{m,n}$. In particular, some of the fusion rules
arise because of zeros of structure constants corresponding to the degenerate primary fields.
We note that not all structure constants, which should vanish
according to MM fusion rules, do vanish in GMM. Thus, GMM can not be considered as a
direct generalization of Minimal Models.

\paragraph{Liouville Field Theory.}

Liouville theory is an irrational CFT with continuous spectrum of
primary fields and central charge
\begin{equation*}
c_L = 1 + 6 Q^2\;.
\end{equation*}
The primary field $V_a$, labelled  by complex parameter $a$,
has conformal dimension 
\begin{equation*}
\Delta^L_a = a(Q - a)\;.
\end{equation*}
Fields with $a = Q/2+iP$ represent the spectrum of the Liouville CFT. The real parameter $P$ is
known as a momentum parameter.

Degenerate fields (i.e. fields, whose Verma module is reducible) are
$V_{m, n} = V_{a_{m, n}} \sim V_{Q - a_{m, n}}$, where 
\begin{equation*}
a_{m,n} = Q/2 - \lambda_{m, n}\;.
\end{equation*}
For theses values of the parameter $a$, a singular vector arises on the $mn$-level of the corresponding Verma module 
\cite{Kac}.

The basic Liouville operator product expansion~\cite{Teschner}
(for the sake of brevity we write $\Delta=\Delta_{Q/2+iP}$ and
$\Delta_{i}=\Delta_{a_{i}}$)
\begin{equation} \label{LOPE}
\begin{aligned} 
\  & V_{a_{1}}(x)V_{a_{2}}(0)=\\
& \ \ \ \ \int' \frac{dP}{4\pi}\left(  x\bar x\right)
^{\Delta-\Delta _{1}-\Delta_{2}}
\mathbb{C}_{a_{1},a_{2}}^{Q/2+iP}\left[
V_{Q/2+iP}(0)\right]  \;,
\end{aligned}
\end{equation}
where the basic structure
constants $\CC_{a_{1}a_{2}}^{Q/2+iP} = \CC^L(a_{1}, a_{2}, Q/2-iP)$~\cite{DO1, LFT}
 (derived from the crossing symmetry in~\cite{Teschner3pt})
have the explicit form (here $a$ denotes $a_{1}+a_{2}+a_{3}$)
\begin{equation} \label{C3}
\CC^L(a_{1}, a_{2}, a_{3})=\left(\!\pi\mu\gamma\!
(b^2)b^{2-2 b^2}\right)^{\!\!(Q-a)/b}\!
\frac{\Upsilon_b(b)}{\Upsilon_b(a-Q)}
\prod\limits_{i=1}^{3}\frac{\Upsilon_b(2 a_i)}{\Upsilon_b(a-2a_i)}\;
,
\end{equation}
where $\Upsilon_b$ is the same ``upsilon'' function as the one, which appears
in the expression for GMM structure constants (see~\cite{DO,LFT}).

The OPE \eqref{LOPE} is continuous and involves integration over the
``momentum'' $P$. The prime on the integral indicates
possible discrete terms, which we discuss in more details in section~\ref{sec:discrete}.
In our computations such extra terms do appear and give an important contribution.
We note that because Liouville theory is non-rational
and fields of interest do not belong to the spectrum, one cannot use OPE literally as in
minimal models but have to apply  instead analytic continuation. For example, if one computes
3-point function with the naive OPE, one often gets zero, which is inconsistent
 with DOZZ formula \eqref{C3} and the 4-point function is inconsistent
with conformal bootstrap.

One can notice, that formulae for the central charge and for the conformal dimensions
in Liouville theory can be obtained from the ones of GMM by $b \to i b$. However
 these two theories are not analytic continuations of each other, because
structure constants of these theories
can not be obtained as analytic continuations. It becomes clear after noticing that
$\Upsilon_b(x)$ function have a natural bound of analyticity with respect to parameter $b$~\cite{Zamol3pt}.

\section{Minimal Liouville Gravity} \label{sec:lg}
In this section we discuss the Minimal
Liouville Gravity correlation numbers on a sphere.

In the framework of the so-called DDK approach~\cite{D,DK}, LG is
a tensor product of the conformal matter (M), represented by
ordinaty or generalized Minimal Models,
Liouville theory (L), and the ghost system (G)
\begin{equation*}
A_{\text{LG}}=A_{\text{M}}+A_{\text{L}}+A_{\text{G}}\;,
\end{equation*}
with the interaction via the construction of the physical fields and the relation for the central
charge parameters
\begin{equation} \label{totc}
c_{\text{M}}+c_{\text{L}}+c_{\text{G}}=0 \;.
\end{equation}

The ghost system (see, e.g.,~\cite{Polchinski1,Verlinde,Fridan}),
consisting of two anticommuting fields $(b,c)$ of spins $(2,-1)$, is the conformal field theory
 with central
charge $c_{\text{G}}=-26$. 

Because of the condition~\eqref{totc},
 we should take (generalized) Minimal Models and LFT with the same value of the parameter $b$.

\paragraph{Physical Fields and Correlation Numbers.}

The physical fields form a space of cohomology classes with respect
to the nilpotent BRST charge $\mathbb{Q}$,
\begin{equation} \label{Q}
\mathbb{Q}=\sum_m{:}\bigg[L_m^{\text{M+L}}+\frac{1}{2}L^{\text{g}}_m\bigg]c_{-m}{:}
-c_0\;.
\end{equation}
In general, we deal with the correlators of fields of the type
\begin{equation} \label{U}
\mathbb{U}_{a}(z,\bar z)=\Phi_{a-b}(z,\bar z)V_{a}(z,\bar z)\;,
\end{equation}
where the choice of the parameters ensures that $\mathbb{U}_{a}(z,\bar z)$ is a $(1, 1)$-form, and
\begin{equation} 
\mathbb{W}_{a}(z,\bar z)= C(z)\bar C(\bar
z)\cdot \mathbb{U}_{a}(z,\bar z)\;, \label{W}
\end{equation}
which is a scalar field. The parameter $a$ can take generic values.

 The $n$-point correlation number on a sphere for these
observables~\cite{BZ1} is
\begin{equation} \label{corrMLG}
I_n(a_1,\ldots,a_n)=\int d^2 z_i \biggl\langle \prod_{i=4}^{n}\mathbb{U}_{a_i}(z_i)
\mathbb{W}_{a_{3}}(z_{3})\,\mathbb{W}_{a_{2}}(z_{2})\,
\mathbb{W}_{a_{1}}(z_{1})\biggr\rangle\;.
\end{equation}
In what follows we shall be interested in correlators of physical fields in the 
minimal gravity constructed from the degenerate matter fields $\UU_{m, n} = \Phi_{m, n} V_{m, -n}$.

For degenerate matter fields there exists an additional set of so-called 
ground ring states~\cite{KlePolGr, WittenGr, BZ1}
\begin{equation}
\mathbb{O}_{m,n}(z,\bar z)=\bar H_{m,n}H_{m,n}\Phi_{m,n}(z,\bar
z)V_{m,n}(z,\bar z)\;.
\end{equation}
The operators $H_{m,n}$ are composed of Virasoro
generators in all three theories and are defined uniquely modulo $\mathbb{Q}$ exact terms. Moreover, if
we introduce the logarithmic counterparts of the ground ring states
$\mathbb O_{m,n}$,
\begin{equation*} 
\mathbb{O}'_{m,n}=\bar H_{m,n}H_{m,n}\Phi_{m,n}V_{m,n}'\;,
\end{equation*}
then we have the following important relation~\cite{HEM},~\cite{BZ1}
\begin{equation} \label{GGU1}
\mathbb{U}_{m,-n}=
B_{m,n}^{-1}\bar\partial\partial \mathbb{O}'_{m,n}\mod \mathbb{Q}\;,
\end{equation}
where $B_{m,n}$ are the coefficients arising in the higher equations
of motion of LFT~\cite{HEM}. For four points,
relation~\eqref{GGU1} allows to reduce the moduli integral in~\eqref{corrMLG}
 to the  boundary integrals if one
of the fields is degenerate, that is $a_i=a_{m,-n}$. It gives 
\begin{equation} \label{4point}
I_4(a_{m,-n},a_2,a_3,a_4)=\kappa
N(a_{m,-n})\left(\prod_{i=2}^4 N(a_i)\right) \Sigma^{(m,n)}(a_2, a_3, a_4)\;,
\end{equation}
where
\begin{equation} \label{SigmaHEM}
\Sigma^{(m,n)}(a)=-m n \lambda_{m, n} + \sum\limits_{i=1}^3
\sum\limits_{r,s}^{(m,n)}|\lambda_i-\lambda_{r,s}|_{\text{Re}}\;, 
\end{equation}
$\lambda_i = Q/2-a_i$ are ``momentum parameters'' and the fusion set
 is $(m,n)=\{1-m:2:m-1,1-n:2: n-1\}$. The first coefficient in~\eqref{4point} is
\begin{equation}\label{kappa}
\kappa=-(b^{-2}+1)b^{-3}(b^{-2}-1) Z_L, \; Z_L = 
\left[ \pi\mu\gamma(b^2)\right]^{Q/b} \frac{1-b^2}{\pi^3 Q \gamma(b^2)\gamma(b^{-2})}
\end{equation}
and the ``leg'' factors  are
\begin{equation*}
N(a)=\frac{\pi}{(\pi \mu)^{(a/b)}}\biggl[\frac{\gamma(2ab-b^2)\gamma(2ab^{-1}-b^{-2})}
{\gamma^{2a/b-1}(b^2)\gamma(2-b^{-2})} \biggr]^{1/2}. \label{N}
\end{equation*}
The expression~\eqref{4point} was derived under the assumption that the number of conformal blocks in
the expansion of the matter sector correlation function is maximally possible, i.e. the number of conformal blocks = $m n$. 
We discuss this point in more details in sections~\ref{sec:GHEM},~\ref{sec:comp}.

 In what follows, we focus on the four-point
correlators in the Lee-Yang series, i.e. 
the parameters $a_i = a_{1, -n_i}$ and $b = \sqrt{2/p}$,
 $\mathcal{I}_4(n_1, n_2, n_3, n_4) = I_4(a_{1, -n_1}, a_{1, -n_2}, a_{1, -n_3}, a_{1, -n_4})$.
\begin{equation}
\mathcal{I}_{4}(n_i)= \int\limits_{\mathcal{M}_{0, 3}} d^2 z \langle 
\mathbb{U}_{1, n_1}(z) \mathbb{W}_{1, n_2}(0)\, \mathbb{W}_{1, n_3}
(1)\,\mathbb{W}_{1, n_4} (\infty) \rangle.
\end{equation}
Taking into account the explicit form of the correlation functions in the ghost sector
\begin{equation*}
\langle C(0) C(1) C(\infty) \rangle=1\;,
\end{equation*}
we obtain
\begin{equation} \label{I41}
\begin{aligned}
\mathcal{I}_{4}(n_i)= \int\limits_{\mathcal{M}_{0, 3}} d^2 z \langle 
\Phi_{1, n_1}(z) \Phi_{1, n_2}(0)\, \Phi_{1, n_3}
(1)\,\Phi_{1, n_4} (\infty) \rangle \times \\
\times \langle 
V_{1, -n_1}(z) V_{1, -n_2}(0)\, V_{1, -n_3}
(1)\,V_{1, -n_4} (\infty) \rangle\;. 
\end{aligned}
\end{equation}
 For further purposes this expression can be conveniently written in more explicit form,
for details, see Appendix~\ref{sec:direct}.

\section{Discrete terms} \label{sec:discrete}

\begin{figure}[h!]
\centering
\def\svgwidth{12cm}
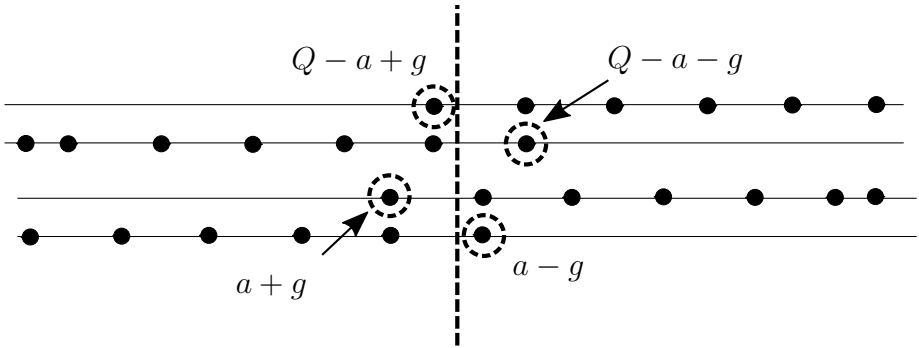
\caption{Poles of structure the constant and discrete terms.}
\label{fig:poles291}
\end{figure}

In the moduli integral~\eqref{I41} the integrand contains four-point
Liouville correlation function. Its computation involves integration in
 momentum $P$, as in~\eqref{LOPE}. The integrand is a product of LFT structure constants~\eqref{C3}
\begin{equation} \label{strprod}
 \CC^L(a_1, a_2, p) \CC^L(Q-p, a_3, a_4)
\end{equation} 
 and the conformal blocks.
 In the case,
where $\Re (|Q/2-a_i|) + \Re (|Q/2-a_j|) < Q/2$ for $i \ne j$
 the contour of integration goes along the real axis. This corresponds to
the fact that in this case the correlator is a sum over intermediate states in the
Hilbert space of Liouville theory. When this condition is not
satisfied, meromorphic continuation of the correlation functions is required.
It can be achieved by deforming the integration contour (see, e.g. ~\cite{Teschner, BZ1}).
Basically, in this case poles of structure constants intersect the real line and
one needs to add corresponding residues to the total integral, as depicted in figure
~\ref{fig:poles291}. These residues are called discrete terms. 
If, for example, $Q/2-a_i > 0$ then the corresponding poles come from zeros
in
 
\begin{multline*} 
\Upsilon(a_i + a_j - p)=\Upsilon(p + (Q/2 - a_i + Q/2 - a_j)), \\
 \Upsilon(a_i + a_j + p - Q) =
\Upsilon(p - (Q/2 - a_i + Q/2 - a_j))\;.
\end{multline*}
In this case one can easily see that the corresponding residues are to be taken at 
\begin{equation*} 
p = (Q/2 - a_i + Q/2 - a_j) - r/b - s b, \; p > Q/2
\end{equation*} 
and in the reflected positions $Q-p$ with the same residues.

We note that expression~\eqref{strprod} in principle may have a second order pole if both of the 
structure constants have poles for the momentum $p$. In what follows we assume that this 
is not the case, then the residues are
 computed easily using quasiperiodicity of $\Upsilon$-function
and the fact that $\Upsilon(\varepsilon) = \Upsilon(b) \varepsilon + O(\varepsilon^2)$.

\section{Generalization of HEM formula} \label{sec:GHEM}

HEM formula
~\eqref{4point},~\eqref{SigmaHEM} was derived assuming that
the fields labelled by 2, 3, 4 are constructed from generic non-degenerate matter fields.
It was assumed that the formula is correct in more general case,
where the number of conformal blocks in the matter sector is equal to
$mn$.

In Minimal Models one can compute four-point function 
$\langle\Phi_{m, n}\Phi_{m_2, n_2}\Phi_{m_3, n_3}
\Phi_{m_4, n_4} \rangle$
using operator product expansions $\Phi_{m, n}\Phi_2 = \sum_{r, s} [\Phi_{r, s}]$ and
 $\Phi_3\Phi_4 = \sum_{r', s'} [\Phi_{r', s'}]$, where the summation goes
 according to MM fusion rules
~\eqref{MMOPE}. The holomorphic contribution
of the pair $(r, s)$ in the intermediate channel defines a conformal block~\cite{BPZ}.
 The number of terms in each 
OPE is not greater then the minimum of products of indices of the fields. Therefore the number
of conformal blocks is not greater then the minimum of $m_i n_i$. For example, in the correlation
function $\langle \Phi_{1, 2}\Phi_{1, 2}\Phi_{1, 2}\Phi_{1, 4}\rangle$
the only field, which can appear in the intermediate channel is
 $\Phi_{1, 3}$ so that the number of conformal blocks 
is less then $2$. Thus the assumption, under which HEM formula was derived, is not
satisfied. However, as we will see later, this formula is correct in more general case.
 The applicability condition is defined by both number of conformal blocks and 
the presence of discrete terms in the Liouville four-point function. In the example
given above
the old formula is correct for $\MM(2/9)$, but is not correct for $\MM(2/11)$.

Most of the consideration below is applicable to the general $\MM(p'/p)$ case, 
though we are going to use it for
the Lee-Yang series.
Our modification of the HEM formula reads:
\begin{equation}
\Sigma_{MHEM} = \Sigma_{HEM} - \sum\limits_{i=2}^4
\sum\limits_{(r,s)\in F_i \cap R_i} 2 \lambda_{r,s}\;, \label{SigmaHEMmod}
\end{equation}
where $F_i$ is the fusion set of $\Phi_{m, n} \Phi_i$ ($\Phi_{m, n} \Phi_i\to\Phi_{r, s}$)
and $R_i$ is the set of discrete terms in the OPE of $V_j V_k$ ($V_j V_k \to V_{r, -s}$) and
$\{i, j, k \} = \{2, 3, 4\}$.

\textbf{Proof of the modified HEM formula.} It appears that for Generalized Minimal 
Models correlation number of four degenerate fields is not well defined.  To compute the correlator
 of degenerate fields we start with the 
correlator of one degenerate and three arbitrary nondegenerate fields $\WW_{a_i}$.
 Then HEM formula reads:
\begin{equation} \label{limcorr}
\langle \UU_{a_{m_1, n_1}} \WW_{a_2} \WW_{a_3} \WW_{a_4} \rangle = 2mn \lambda_{m, n}
 + \sum\limits_{i=2}^4
\sum\limits_{r,s}^{(m,n)} \left( |\lambda_i-\lambda_{r,s}|_{\text{Re}} - \lambda_{m, n} \right).
\end{equation}
We are interested in the limit $a_i \to a_{m_i, n_i}$.
The direct approach based on the conformal block decomposition gives the formula for 
the correlator~\eqref{limcorr}:

\begin{multline} \label{limcorrexpl}
\mathcal{I}_4(a_{m_1, n_1}, a_i)= 
2 \sum_{k} \CC^k_{12} \CC_{k34} \\ \pi^2 \int'
\frac{dP}{4\pi}  \CC^{L}(a_{m_1, n_1}, a_2, Q/2-iP) \CC^{L}(Q/2+iP, a_3, a_4) \\
\sum_{l} \sum_{j}  \Big( b_j(P) b_{l-j}(P) 
 \Phi(\Delta^L(P)+\Delta^M_k-1, j, l-j) \Big) + \circlearrowright\,\,,
\end{multline}
where $\circlearrowright$ denotes two other terms with cyclic permutations of $a_2, a_3, a_4$
(see Appendix~\ref{sec:direct} for details). Let us focus on
 one of these terms. In~\eqref{limcorrexpl}
some Minimal Model structure constants $\CC^k_{12}$ and $\CC_{k34}$ become zero
in the desired limit\footnote{
There are some complications when these structure constants do not vanish even if they 
should do according to the fusion rules~\cite{Zamol3pt}, but this is not the case for
 Lee-Yang series and 
 is not discussed here.}.
Let us denote the corresponding terms in~\eqref{limcorrexpl} as
 \begin{equation} \label{extraterms}
 \langle \UU_{a_{m_1, n_1}}  \WW_{a_{m_2, n_2}} \WW_{a_{m_3, n_3}} \WW_{a_{m_4, n_4}}
 \rangle^k + \circlearrowright.
\end{equation}
 When the matter is represented by Minimal Models, these terms do not appear in
 the expression for the correlation numbers because of the fusion rules.
However, sometimes these terms
do not vanish automatically in the limit, so that to get an answer for Minimal Models
we take the limit of~\eqref{limcorr} and then subtruct terms~\eqref{extraterms}
\begin{multline} \label{corrform}
\langle \UU_{a_{m_1, n_1}} \WW_{a_{m_2, n_2}} \WW_{a_{m_3, n_3}} \WW_{a_{m_4, n_4}} \rangle = \\
 \lim_{a_i\to a_{m_i, n_i}} \left[\langle \UU_{a_{m_1, n_1}} \WW_{a_2} \WW_{a_3} \WW_{a_4} \rangle
\right]
 - \left(\sum_{k}
\langle \UU_{a_{m_1, n_1}}  \WW_{a_{m_2, n_2}} \WW_{a_{m_3, n_3}} \WW_{a_{m_4, n_4}}
 \rangle^k 
 + \circlearrowright\right)\;.
\end{multline}

Let us compute the contribution of~\eqref{extraterms}.
 Some of these terms do not vanish because $\Phi(\Delta^L(P)+\Delta^M_k-1, j, l-j)$, arising from
the $x$-integration, has a pole and annihilates zero appearing in the structure constant~\eqref{mat3pt}.
Explicitly one has (see~\eqref{phi})
\begin{equation*} 
\Phi(A,r,l)=\frac{(16)^{2A}}{\pi(2A+r+l)}\int_{-1/2}^{1/2}\cos(\pi(r-l)x)e^{-\pi
\sqrt{1-x^{2}}\left(  2A+r+l\right)  }dx\;,
\end{equation*}
so that it has a pole when $2A+r+l = 0$ and $r-l$ is odd or zero. In our case 
it implies $r=l=0$ and $A=0$,
 which leads us to the conclusion that in the intermediate channel the Liouville
 dimension $\Delta^L(P)$ should be dressing
for the matter dimension $\Delta^M_k$ in the sense that $\Delta^L(P)+\Delta^M_k=1$.
 This can be possible only if Liouville correlation function
has specific discrete terms, i.e. 
$i P \to \lambda_{m, -n}$.  The first thing to notice is that nonzero terms~\eqref{extraterms}
appear precisely if $k = (r, s) \in F_i \cap R_i$ as in the formula~\eqref{SigmaHEMmod}.

 Let us compute the value of each of these terms.
We have 
\begin{multline} \label{terms}
\lim_{a_i\to a_{m_i, n_i}}\langle \UU_{a_{m_1, n_1}}  \WW_{a_2} \WW_{a_3} \WW_{a_4} \rangle^k_F = 
\CC^{G, p}_{a_{m_1, n_1}, a_{m_2, n_2}} \CC^G _{p, a_{m_3, n_3}, a_{m_4, n_4}} = \\
 \CC^{G}_{a_{m_1, n_1}, a_{m_2, n_2}, p}
(D^G_{p, p})^{-1} \CC^G _{p, a_{m_3, n_3}, a_{m_4, n_4}} = 2 \lambda_k \kappa
 \prod_{j=1}^4 N(a_{m_i, n_i})\;,
\end{multline}
 where $D^G$ and $\CC^G$ denote MLG two- and
three-point functions, $\kappa$ is given in~\eqref{kappa} and $\lambda_k$ is
$Q/2-a_k$. Taking~\eqref{corrform},~\eqref{terms} into account one derives modified
 HEM formula~\eqref{SigmaHEMmod}.

Let us now accurately prove~\eqref{terms}. We start from the formula~\eqref{limcorrexpl}.
Taking residue in the discrete terms and using reflection relation in LFT,
$\CC^{L}(a_{m_1, n_1}, a_2, p_k) R_L(p_k)^{-1} = \CC^{L}(a_{m_1, n_1}, a_2, Q-p_k)$, we have
\begin{multline} \label{limcorrexpl1}
 \langle \UU_{a_{m_1, n_1}}  \WW_{a_2} \WW_{a_3} \WW_{a_4} \rangle^k_F= 
2 \,\CC^M(\alpha_{m_1, n_1}, \alpha_2, \alpha_k) \CC^M(\alpha_k, \alpha_3, \alpha_4) \\  
\pi^2  \CC^{L}(a_{m_1, n_1}, a_2, p_k) R_L(p_k)^{-1} \Res_{p \to p_k}
 [\CC^{L}(p_k, a_3, a_4)] \\
\sum_{l} \sum_{j}  \Big( b_j(p_k) b_{l-j}(p_k) 
 \Phi(A_k(p_k), j, l-j) \Big) + \circlearrowright \;.
\end{multline}
In the last formula we used the notations $p = Q/2 + iP$, $A_k(P) = \Delta^L_{a_k} +\Delta^M_{\alpha_k}-1$ and $p_k$ is the value of $p$ 
corresponding to the discrete term of interest.
 In~\eqref{limcorrexpl1} we also took into account two equaivalent symmetric residues,
 which produces the factor of 2. Now we denote $\varepsilon = p_k - a_k$, 
where $\Delta^L_{a_k} +\Delta^M_{\alpha_k}=1$ and $\Delta^M_{\alpha_k}$ is the dimension
of the intermediate field in the MM conformal block. We ignore 
terms of order $o(\varepsilon)$ and multiply 
Minimal Model structure constants by Liouville ones to get MLG three-point functions.
In this way we obtain
\begin{multline} \label{limcorrexpl2}
 \langle \UU_{a_{m_1, n_1}}  \WW_{a_2} \WW_{a_3} \WW_{a_4} \rangle^k_{\mathbf{F}} \sim
2 \,\CC^M(\alpha_{m_1, n_1}, \alpha_2, \alpha_k) \cdot \CC^{L}(a_{m_1, n_1}, a_2, p_k) \\
\CC^M(\alpha_k, \alpha_3, \alpha_4) \cdot \, (-\varepsilon) \CC^{L}(\alpha_k, a_3, a_4) \\
\pi^2   R_L(p_k)^{-1}
\sum_{l} \sum_{j}  \Big( b_j(p_k) b_{l-j}(p_k) 
 \Phi(A_k(p_k), j, l-j) \Big) + \circlearrowright\;.
\end{multline}
To compute this expression we note that $\Phi$ has a pole
in $\varepsilon$ only if $j=l=0$, so that
 we can ignore other terms. Using the explicit formula
 for $\Phi$ we find
\begin{equation} \label{philim}
\Phi(A_k(p_k), 0, 0) \sim \frac{1}{2 \pi A_k(p_k)} \sim \frac{1}{2 \pi \Delta^L(p_k)' \varepsilon}.
\end{equation}
Now we expand the value of $R_L(p_k)$:
\begin{equation}\label{RL}
R_L(a) = (\pi \mu \gamma(b^2))^{(Q-2a)/b}\frac{\gamma(2ab-b^2)}{b^2 \gamma(2-2ab^{-1}+b^{-2})}
\end{equation}
and two- and three-point functions in MLG are correspondingly:
\begin{equation}
\begin{aligned}
&D^G_{a, a} = \frac{\kappa}{2 \lambda_a} N(a)^2,\\
&\CC^G_{a_1, a_2, a_3} = b\kappa \prod_{i=1}^3{N(a_i)}.
\end{aligned}
\end{equation}
Using these expressions we finally arrive to the formula~\eqref{terms} and thus
 prove~\eqref{SigmaHEMmod}.

\hfill\ensuremath{\square}  \\

For Lee-Yang series we can further simplify~\eqref{SigmaHEMmod}. Without loss of generality let
 $n_1\le n_2\le n_3\le n_4\le s, \; p = 2s+1$. Then   only the term
\begin{equation} \label{LYDT}
\sum_{(1, s) \in R_4} 2\lambda_{1, s}
\end{equation}
survives in the sum. If $\sum_i n_i$ is even, then the last expression is equal to
\begin{equation*}
\sum_{s = n_2+n_3+1 \; : \; 2}^{\min(n_1+n_4-1, s)} 2\lambda_{1, s} = \frac{1}{2\sqrt{2p}}
\left(\hat{F}(\min(n_1+n_4, n_2+n_3))- \hat{F}(n_1+n_4)\right),
\end{equation*}
where $\hat{F}(n) = (s+1-n)(s-n)\theta(n\le s)$.
If $\sum_i n_i$ is odd, then~\eqref{LYDT} equals to
\begin{equation*}
\sum_{s = n_2+n_3+1 \; : \; 2}^{s} 2\lambda_{1, s} = \frac{1}{2\sqrt{2p}}\hat{F}(n_2+n_3) =
 \frac{1}{2\sqrt{2p}}\left(\hat{F}(\min(n_1+n_4, n_2+n_3))- \hat{F}(n_1+n_4)\right),
\end{equation*}
where the last equality is due to $n_1+n_4 > s$ and $n_2+n_3 < s$.

Now for Lee-Yang series we can rewrite~\eqref{SigmaHEMmod} as

\begin{equation} \label{SigmaHEMmodLY}
\Sigma_{MHEM} = \Sigma_{HEM} - \frac{1}{2\sqrt{2p}}\left(\hat{F}(\min(n_1+n_4, n_2+n_3))-
 \hat{F}(n_1+n_4)\right). 
\end{equation}

\section{Comparison with Douglas equation approach} \label{sec:comp}

In this section we compare our results with the results of the Douglas equation approach.

Using identification $\Phi_{1, n} = \Phi_{1, p-n}$ in Lee-Yang series we will study fields
 $\UU_{1, n}$
with $n \le s$, where $p = 2s+1$. Our modified HEM approach gives formula~\eqref{SigmaHEMmodLY}.
 For comparison purposes we 
consider a normalization independent version of this formula:

\begin{multline} \label{HEMformula}
\frac{\left<\left< \UU_{m_1, n_1} \UU_{m_2, n_2} \UU_{m_3, n_3} \UU_{m_4, n_4} \right>\right>}
{\left(\prod^{4}_{i=1} \left<\left< \UU^2_{m_i, n_i} \right>\right>\right)^{1/2}} = \\
\frac{\prod^{4}_{i=1} |m_i p - n_i p'|^{1/2} }{2p(p+p')(p-p')}\Big( \sum\limits^{4}_{i=2}
\sum\limits^{m_1-1}_{r = -(m_1-1)}\sum\limits^{n_1-1}_{t = -(n_1-1)} |(m_i -r)p - (n_i-t)p'| - 
m_1 n_1 (m_1 p + n_1 p') \Big) = \\
\frac{\prod^{4}_{i=1} |m_i p - n_i p'|^{1/2} }{2p(p+p')(p-p')} (- 2\sqrt{p p'} \Sigma_{MHEM} (m_i, n_i))\;,
\end{multline}
where $p'=2$ and $n_i=1$.\footnote{We do not specify $p'$ and $m_i$ in~\eqref{HEMformula} in order to
make the structure of this formula more clear.}
We denote $\Sigma'(m_i, n_i) = - 2\sqrt{p p'} \Sigma (m_i, n_i)$ and expect it to be an 
integer number, so that in the comparison it will be the most convenient quantity.

The numerical quantity to be compared with $\Sigma'_{MHEM}$ is
\begin{equation} \label{sigmaHEM2}
\Sigma'_{NUM}(m_i, n_i) = - 2\sqrt{p p'} \frac{I_4(m_i, n_i)}{\prod^4_{i=1} N(m_i, n_i) \kappa}\;.
\end{equation}

In the framework of the Douglas equation approach there are two formulae for the
four-point correlation numbers.
 First of them~\cite{BZ_MM, BDM} after renormalization can be written as
\begin{multline} \label{SigmaDSE}
\Sigma'_{DSE}(n_i) = -\hat{F}(0) + \sum^{4}_{i=1}\hat{F}(n_i) \\
 -\hat{F}(\mathrm{min}(n_1+n_2,n_3+n_4)) 
-\hat{F}(\mathrm{min}(n_1+n_3,n_2+n_4))-\hat{F}(\mathrm{min}(n_1+n_4,n_3+n_2))\;,
\end{multline}
where $\hat{F}(n) = (s+1-n)(s-n) \theta(n \le s)$.

The second one is proposed in~\cite{BelavinRud}
and coincides with the above one when the number of conformal blocks is maximal and
does not otherwise.

\begin{prop}  \label{formeq}
The formula for four-point correlation numbers in Douglas equation approach 
is equivalent to the modified HEM formula:
\begin{equation*}
\Sigma_{MHEM}(n_i) = \Sigma_{DSE}(n_i)\;.
\end{equation*}
\end{prop}
Moreover, if there are no discrete terms in the operator product expansion
 $V_{1, n_2} V_{1, n_3}$, then we also have $\Sigma_{HEM}=\Sigma_{MHEM}$.
The proof can be found in Appendix~\ref{sec:formcheck}.

All our numerical computations of correlation numbers in various models
confirm that the formula~\eqref{SigmaHEMmodLY} is correct.
In order to give some reference points we list some of the numerical results
compared with Douglas equation approach and with the old HEM formula in
 tables~\ref{table:t1},\ref{table:t4}.  In the tables correlator
 $12 \; 12 \; 12\; 14$ means $\langle \UU_{1, 2} \WW_{1, 2}
 \WW_{1, 2}\WW_{1, 4} \rangle$ and so on. Sign * after the correlator means that
 there are discrete
terms in Liouville OPE of any of the four fields, sign $^\dagger$ means
that there is a discrepancy between different approaches ($\Sigma'_{NUM, DSE, HEM}$ 
correspond to numerical computation, Douglas equation approach and higher equations of
motion approach respectively).

In the table~\ref{table:t1} we give some
results on correlation numbers in different models. Note that in the table we also
presented the results for the Minimal Model $\MM(4/15)$, which does not belong to the Lee-Yang series.
 We list a larger set of correlation numbers in the table~\ref{table:t4} for the model $\MM(2/15)$.

\section{Discussion} 
\label{sec:disc}

We have considered the direct approach to Liouville Minimal Gravity.
 Our main result is the formula~\eqref{SigmaHEMmod} for four-point
correlation numbers in the Lee-Yang series. This 
formula generalizes the old one~\eqref{SigmaHEM} proposed in~\cite{BZ1}. We show that
our modified HEM formula is equivalent to the DSE formula~\eqref{SigmaDSE}.
We also performed numerical checks, which
 confirm our results in the region of parameters where the old
formula was not applicable.

Below we state some questions which naturally arise from the present considerations.

 If the matter sector is
represented by the Minimal Model with $p'>2$, in Douglas equation approach 
it is impossible to fulfil all the Minimal Model fusion rules as was shown in~\cite{BDM, unitary}. 
So it would be interesting to see 
how does the correspondence between DSE and conformal field theory approaches
extends to other Minimal Models. 

 In~\cite{Zamol3pt} there was obtained a formula
for three-point functions in GMM. It coincides with the one obtained by Dotsenko and Fateev
 in~\cite{DF1} when it is not forbidden by fusion rules. But for some reason this formula gives a nonzero result
for some structure constants which should vanish according to the fusion rules.
Taking into account this fact would clearly lead to further complications for four-point correlation numbers
in general Minimal Models, as mentioned in Section~\ref{sec:GHEM}.
In~\cite{Zamol3pt} the prescription to obtain MM from GMM
 is to multiply the GMM structure constants by
fusion algebra constants.
 As far as we know, there is no good understanding of this phenomenon, but it
can also be connected with the previous question and with the fusion rules problem in MLG.
For instance, without this additional restriction MLG three-point functions are always nonzero, that
requires a better understanding.
 Some insight to this problem can be found in~\cite{Santachiara}, where Liouville theory
with $c\le1$ (GMM in our language) is discussed.\\

\begin{center}%
\begin{table}[bp] \centering
\begin{tabular}
[c]{|c|c|c|c|}\hline $m_i n_i$ & $|\Sigma'_{NUM}(m_i, n_i)|/2$ num.
 & $\Sigma'_{DSE}(m_i, n_i)/2$ exact & $\Sigma'_{HEM}(m_i, n_i)/2$ exact \\
\hline
 2/9 & - & - & -\\
12 12 12 12 & 2.00002 & 2 & 2 \\
13 13 12 12 &  2.00031 & -2 & -2 \\
12 14 12 12 &  1.00003 & -1 & -1 \\
13 12 13 13 &  4.00016 & -4 & -4 \\
 2/11 & - & - & -\\
13 15 13 13 & 5.99976 & -6 & -6 \\
12 14 12 12*$^\dagger$ & 1.000001  & 1 & 2 \\
 2/13 & - & - & -\\
12 14 12 12*$^\dagger$ & 3.0001 & 3 & 6 \\
 4/15 & - & - & -\\
13 17 13 13 &  2.00009 & N/A & -2 \\
13 15 13 13 & 10.9998 & N/A & -11\\ \hline
\end{tabular}
\caption{Numerical data for $\Sigma'$. * - means discrete terms. $^\dagger$ - discrepancies.}
\label{table:t1}
\end{table}%
\end{center}

\begin{center}%
\begin{table}[tbp] \centering
\begin{tabular}
[c]{|c|c|c|c|}\hline $m_i n_i$ & $|\Sigma'_{NUM}(m_i, n_i)|/2$ num.
 & $\Sigma'_{DSE}(m_i, n_i)/2$ exact & $\Sigma'_{HEM}(m_i, n_i)/2$ exact \\
\hline
 2/15 & - & - & -\\
12 12 13 15*$^\dagger$ & 3.08 & 3 & 6 \\
12 12 14 16*$^\dagger$ &  1.025 & 1 & 2 \\
12 12 15 15* &  1.98 & 2 & 2 \\
12 12 15 17* &  1.03 & -1 & -1 \\
12 12 16 16* &  2.06 & -2 & -2 \\
12 12 17 17* &  4.09 & -4 & -4 \\
12 13 13 16*$^\dagger$ &  1.01 & 1 & 2 \\
12 13 14 15* &  1.99 & 2 & 2 \\
12 13 14 17* &  1.015 & 1 & 1 \\
12 13 15 16* &  2.03 & -2 & -2 \\
12 13 16 17* &  5.07 & -5 & -5 \\
12 14 14 14* &  1.995 & 2 & 2 \\
12 14 14 16* &  2.02 & -2 & -2 \\
12 14 15 15* &  2.01 & -2 & -2 \\
12 14 15 17* &  5.04 & -5 & -5 \\
12 14 16 16* &  6.05 & -6 & -6 \\
12 14 17 17* &  8.04 & -8 & -8 \\
13 13 13 15*$^\dagger$ &  1.995 & 2 & 3 \\
13 13 13 17*$^\dagger$ &  0.999 & -1 & 0 \\
13 13 14 14* &  2.93 & 3 & 3 \\
13 13 14 16* &  2.01 & -2 & -2 \\
13 13 15 15* &  3.03 & -3 & -3 \\
13 13 15 17* &  6.05 & -6 & -6 \\
13 13 16 16* &  7.04 & -7 & -7 \\
13 13 17 17* & 9.05 & -9 & -9 \\ \hline
\end{tabular}
\caption{Numerical data for $\Sigma'$. * - means discrete terms. $^\dagger$ - discrepancy. }
\label{table:t4}
\end{table}%
\end{center}

\section*{Acknowledgements}
We thank Alexander Belavin for useful discussions and all other people with
whom authors discussed the subject.
The work was performed 
with financial support from the Russian Science Foundation (Grant No.14-12-01383).
V.B. thanks G. Mussardo and INFN for possibility to visit SISSA, where the final part
of this work was performed.

\appendix
\section{Conformal block decomposition in correlation numbers} \label{sec:direct}
In this appendix we derive convenient representation for the correlation numbers.\\

The following considerations are known in the literature,
see e.g.~\cite{ZamolLeeYang}. We start from the formula~\eqref{I41} for the correlation numbers and
 use the symmetry of the integrals under modular transformations in order to
reduce the integration from the whole complex plane to the fundamental
domain. The modular subgroup of projective transformations divides the
complex plane into six regions. The fundamental region is defined as
$\mathbf{G=}\{\operatorname*{Re}x<1/2;\;\left| 1-x\right| <1\}$. The other
five regions are mapped to the fundamental one using one of the
transformations $\mathcal{A},\mathcal{B},\mathcal{A}\mathcal{B},
\mathcal{B}\mathcal{A},\mathcal{A}\mathcal{B}\mathcal{A}$,
where $\mathcal{A}$: $z\rightarrow 1/z$ and $\mathcal{B}$: $z\rightarrow
1-z$. Combining the projective transformations of the fields and the
corresponding change of the variables in the integrals, we reduce the
integration to the fundamental region. We note that the Jacobian of the
transformation exactly cancels the transformation of the fields
because their total conformal dimension is $1$. Then,
\begin{equation} \label{I4bcomp}
\begin{aligned}
\mathcal{I}_4(n_i)=2\int_{\mathbf{G}} d^2 z \bigg(\langle \WW_1(0)\UU_2(z)\WW_3(1)
\WW_4(\infty)\rangle+
 \langle \WW_3(0)\UU_2(z)\WW_1(1)\WW_4(\infty)\rangle+\\
+\langle \WW_4(0)\UU_2(z)\WW_3(1)\WW_1(\infty)\rangle\bigg),
\end{aligned}
\end{equation}
where the factor 2 in front counts the equivalent projective images 
(the order of the last two fields is not relevant) and $\UU_i, \WW_i$ stand
for $\UU_{a_i}, \WW_{a_i}$.

\paragraph{Conformal block decomposition.}

For a
while, we omit some arguments that are easily reconstructed in the
final expressions.  In the matter sector,
\begin{equation} \label{I4bev2}
\begin{aligned}
&\langle \Phi_1(0) \Phi_2(z)  \Phi_3(1) \Phi_4(\infty) \rangle= \sum_k c^{(1)}_k
 |\FF^{(1)}_k(z)|^2\;,\\
&\langle \Phi_3(0) \Phi_2(z)  \Phi_1(1) \Phi_4(\infty) \rangle= \sum_k c^{(1)}_k
 |\FF^{(3)}_k(z)|^2\;,\\
&\langle \Phi_4(0) \Phi_2(z)  \Phi_3(1) \Phi_1(\infty) \rangle= \sum_k c^{(1)}_k
 |\FF^{(4)}_k(z)|^2\;.
\end{aligned}
\end{equation}
Here the index $k$ corresponds to the channels in the
degenerate OPE of the fields $\Phi_i$ and the
coefficients $c_k$ are related to the basic structure constants
 ~\cite{DF1, Zamol3pt}:
\begin{equation*} 
c^{(1)}_k = \CC^k_{12} \CC^k_{34}, c^{(3)}_k = \CC^k_{32} \CC^k_{14},
 c^{(4)}_k = \CC^k_{42} \CC^k_{31}\;.
\end{equation*} 
In~\eqref{I4bev2}, $\FF^{(i)}_k$ denotes the conformal blocks appearing in the
$k$-channel for the given correlation function. 
In the Liouville sector we have
\begin{equation} \label{I42bev2}
\begin{aligned} 
&\langle V_1(z) V_2(0) V_3(1) V_4(\infty) \rangle= \mathcal{R}\int'
\frac{dP}{4\pi} r^{(1)}(P) |\FF^{(1)}(P,z)|^2\;, \\
&\langle V_3(z) V_2(0) V_1(1) V_4(\infty) \rangle= \mathcal{R}\int'
\frac{dP}{4\pi} r^{(3)}(P) |\FF^{(3)}(P,z)|^2\;,  \\
&\langle V_4(z) V_2(0) V_3(1) V_1(\infty) \rangle= \mathcal{R}\int'
\frac{dP}{4\pi} r^{(4)}(P) |\FF^{(4)}(P,z)|^2\;,
\end{aligned}
\end{equation}
where
\begin{equation*}
\mathcal{R} r^{(1)}(P)=
\CC^L(a_1,a_2, Q/2+iP)\CC^L(Q/2-iP, a_3, a_4)
 \end{equation*}
and so on. Here $\mathcal{R}$ stands for the momentum independent part of the product.
 In what follows we omit upper subscripts pointing permutations of the fields
 and summation with respect to them
in the correlators.
\paragraph{The Modular Integral.}

It is efficient~\cite{ellrec} to go to the universal cover of the moduli space 
$\mathcal{M}_{0, 3} = S^2 \backslash \{0, 1, \infty\}$, that is to use elliptic transformation
in the integration. We use the map
\begin{equation*}
\tau=i \frac{K(1-z)}{K(z)}\;,
\end{equation*}
where $K(z)$ is the complete elliptic integral of the first kind
\begin{equation*}
K(z)=\frac{1}{2} \int_0^1 \frac{d t}{y}
\end{equation*}
and $y^2=t(1-t)(1-z t)$. It can be verified that
\begin{equation*}
dz=\pi z(1-z)\theta_3^4(q)d\tau\;,
\end{equation*}
where the elliptic nome parameter
\begin{equation*}
q=e^{i \pi \tau}
\end{equation*}
and theta constant
\begin{equation*}
\theta_3(q)=\sum_{n=-\infty}^{\infty}q^{n^2}\;.
\end{equation*}
Following~\cite{ellrec} we can write
\begin{equation} \label {confel}
\FF(\Delta_i, \Delta | q) = (16 q)^{\Delta_p - \Delta_0} z^{\Delta_0-\Delta_1-\Delta_2}
(1-z)^{\Delta_0-\Delta_2-\Delta_3} \theta_3^{12 \Delta_0 - 4 \sum \Delta_i}(q)
H(\Delta_i, \Delta | q)\;,
\end{equation}
in order to represent integral~\eqref{I4bcomp} in the following form
\begin{align}\label{intready}
\mathcal{I}_4(a_i)= 2 \sum_{i}^{1, 3, 4} \pi^2 \mathcal{R}\int'
\frac{dP}{4\pi} \sum_{k} r^{(i)}(P) c_k^{(i)}
\int_{\mathbf{F}}
|16 q^{A_k(P)} H^{(i)}_k(q | \Delta^L_p) H^{(i)}_k(q)|^2 \dif^2 \tau\;,
\end{align}
where $\mathbf{F}$\textbf{\ }$=\left\{  \left|  \tau\right|
>1;\;\left| \operatorname*{Re}\tau\right|  <1/2\right\}$, $A_k(P) = \Delta^L(P) + \Delta^M_k - 1$
 is sum of conformal dimensions in the intermediate channel minus 1 and
 $H(\Delta_i, \Delta | q)$ is a series in $q$ of the form $1 + O(q)$, which is
computed using recurrence relation \cite{ellrec}.

\paragraph{Numerics.}

With~\eqref{intready}, the calculation reduces to the
numerical integration of several integrals of the general form
\begin{equation} \label{formint}
\int_{\mathbf{F}}|z(1-z)\theta_{3}^{4}(q)\mathcal{F}_P(z)|^2
d^2\tau\;,
\end{equation}
where $\mathcal{F}_P(z)$ is some Liouville conformal block like
in~\eqref{I4bev2} or some more complicated composite expression like
in~\eqref{I42bev2}. The integrand can be developed as a power
series in $q$ according to
\begin{equation} \label{Hr}
z(1-z)\theta_{3}^{4}(q)\mathcal{F}_P(z)=\left( 16q\right)^{\alpha}
\sum_{r=0}^{\infty} b_{r}(P)q^{r}\;
\end{equation}
and the same for $\bar{q}$.
In each term, we
can integrate in $\tau_{2}=\operatorname*{Im}\tau$ explicitly with
the result conviniently represented in terms of the function
\begin{equation} \label{phi}
\Phi(A,r,l)=\int_{\mathbf{F}}d^{2}\tau\left|  16q\right|
^{2A}q^{r}\bar
q^{l}=\frac{(16)^{2A}}{\pi(2A+r+l)}\int_{-1/2}^{1/2}\cos(\pi(r-l)x)e^{-\pi
\sqrt{1-x^{2}}\left(  2A+r+l\right)  }dx\;.
\end{equation}

Using explicit formulae~\eqref{intready},~\eqref{Hr},~\eqref{phi}
we finally obtain the following expression for~\eqref{intready}:

\begin{multline} \label{intnumerics}
\mathcal{I}_4(a_i)= \\
2 \sum_{i}^{1, 3, 4} \sum_{k} c_k^{(i)} \pi^2 \mathcal{R}\int'
\frac{dP}{4\pi}  r^{(i)}(P) \sum_{l} \sum_{j}  \Big( b_j(P) b_{l-j}(P) 
 \Phi(A_k(P), j, l-j) \Big)\;,
\end{multline}
where
$b_j(P) = [q^j](H^{(i)}(q | \Delta^L_p) H^{(i)}_k(q))$ is a 
$q^j$th term in the expansion of the elliptic conformal blocks 
and $A_k(P) = \Delta^L(P) + \Delta^M_k - 1$ as above.
Each term in~\eqref{intnumerics} is suppressed by a factor
$\max_{\mathbf{F}}\left| q\right|  ^{2l}$ and the series in $l$
converges very rapidly. 

Main source of numerical errors in these computations is a method of computing
product of Liouville structure constants, namely functions $r^{(i)}(P)$.
\section{Proof of the proposition~\ref{formeq}} \label{sec:formcheck}

Here we prove that 
\begin{equation}\label{eqty}
\Sigma'_{MHEM}(n_i)=\Sigma'_{DSE}(n_i).
\end{equation}
Let $n_1\le n_2\le n_3\le n_4\le s$.
For our purposes we write $\Sigma'_{DSE}$ and $\Sigma'_{MHEM}$ as
\begin{equation}
\Sigma'_{DSE} = -\hat{F}(0) + \sum_{i=1}^4 \hat{F}(n_i) - \hat{F}(n_1+n_2) - \hat{F}(n_1+n_3)
- \hat{F}(\min(n_1+n_4, \; n_2+n_3))
\end{equation}
and
\begin{equation}
\Sigma'_{MHEM} = \sum_{i=2}^4\sum_{t}^{(n_1)}|p-2(n_i+t)| - n_1 (p+2 n_1) + 
-\hat{F}(\min(n_1+n_4, n_2+n_3)) + \hat{F}(n_1+n_4).
\end{equation}
First, we need to show that 
the old formula $\Sigma'_{HEM}$ coincides with $\Sigma^*_{DSE}$,
 where we have introduced
\begin{equation}
\Sigma^*_{DSE} = -\hat{F}(0) + \sum_{i=1}^4 \hat{F}(n_i) - \hat{F}(n_1+n_2) - \hat{F}(n_1+n_3)
- \hat{F}(n_1+n_4)\;.
\end{equation}
If $n_1+n_i \le s+1$, in $\Sigma'_{HEM}$ all expressions under modules are positive and in 
$\Sigma'_{DSE}$ all $\hat{F}$ are equal to $\hat{F}_0$, so that both $\Sigma'$ simplify to
\begin{equation} \label{smplsig}
2 n_1 (p - \sum_i n_i)\;.
\end{equation}
When $n_1+n_i \ge s+2$ for some $i$, $\Sigma'_{HEM}$ gets a correction
to~\eqref{smplsig} because of the modules equal to 
\begin{equation*}
-2\!\!\!\!\!\!\!\sum_{t\; :\; p-2(n_i+t)<0}(p - 2(n_i+t)) = (s+1-n_1-n_i)(s-n_1-n_i)\;.
\end{equation*}
$\Sigma^*_{DSE}$ gets a correction because of the Heaviside theta function equal to
$\hat{F}(n_1+n_i) = (s+1-n_1-n_i)(s-n_1-n_i)$, which finishes the proof.

Now the initial statement~\eqref{eqty}  follows immediately from the definitions.

\bibliographystyle{JHEP}
\bibliography{mlg}

\end{document}